\documentclass{ifacconf}

\usepackage{graphicx} 
\usepackage{natbib}      

\usepackage{algorithm, algpseudocode}
\usepackage{tabularx}
\usepackage{amsmath}
\usepackage{amssymb}
\usepackage{xcolor}
\usepackage{pgfplots} 
\usepackage{xspace}
\usepackage{changes}

\newcommand{\mrm}[1]{\mathrm{#1}}
\newcommand{\R}{\operatornamewithlimits{{\mathbb{R}}}} 

\newcommand{\diff}{\ensuremath{\mathrm{d}}}

\newcommand{\Dc}{\mathcal{D}}
\newcommand{\Hc}{\mathcal{H}}
{\left[\begin{smallmatrix}}
	{\end{smallmatrix}\right]}%

\newcommand{\mmVal}[3]{$#1\mathrm{\:=\:}#2\,\mathrm{#3}$\xspace}

\begin{document}
	\graphicspath{{figures/}}
	
\begin{frontmatter}

\title{State and parameter estimation for model-based retinal laser treatment\thanksref{footnoteinfo}} 

\thanks[footnoteinfo]{The collaborative project "Temperature controlled retinal laser treatment" is funded by the German Research Foundation (DFG) under the project number 430154635 (MU 3929/3-1, WO 2056/7-1, BR 1349/6-1). MS was also funded by the DFG (grant WO\ 2056/2-1, project number 289034702). 
	KW gratefully acknowledges funding by the German Research Foundation (DFG; grant WO\ 2056/6-1, project number 406141926). This work has been submitted to IFAC for possible publication.}

\author[First]{Viktoria Kleyman}, 
\author[Second]{Manuel Schaller}, 
\author[Second]{Mitsuru Wilson},
\author[Third]{Mario Mordmüller},
\author[Third]{Ralf Brinkmann},
\author[Second]{Karl Worthmann} and
\author[First]{Matthias A. Müller}

\address[First]{Leibniz University Hannover, Institute of Automatic Control, (e-mail: \{kleyman,mueller\}@irt.uni-hannover.de)}
\address[Second]{Technische Universit\"at Ilmemau, Institute for Mathematics, (e-mail: \{manuel.schaller,mitsuru.wilson,karl.worthmann\}@tu-ilmenau.de)}
\address[Third]{University of Lübeck, Institute of Biomedical Optics, (e-mail: \{m.mordmueller,ralf.brinkmann\}@uni-luebeck.de)}

\begin{abstract}  
We present an approach for state and parameter estimation in retinal laser treatment by a novel setup where both measurement and heating is performed by a single laser. In this medical application, the temperature that is induced by the laser in the patient's eye is critical for a successful and safe treatment. To this end, we pursue a model-based approach using a model given by a heat diffusion equation on a cylindrical domain, where the source term is given by the absorbed laser power. The model is parametric in the sense that it involves an absorption coefficient, which depends on the treatment spot and plays a central role in the input-output behavior of the system. After discretization, we apply a particularly suited parametric model order reduction to ensure real-time tractability while retaining parameter dependence. We augment known state estimation techniques, i.e., extended Kalman filtering and moving horizon estimation, with parameter estimation to estimate the absorption coefficient and the current state of the system. Eventually, we show first results for simulated and experimental data from porcine eyes. 
We find that, regarding convergence speed, the moving horizon estimation slightly outperforms the extended Kalman filter on measurement data in terms of parameter and state estimation, however, on simulated data the results are very similar. 
\end{abstract}


\end{frontmatter}

\section{Introduction}

Retinal photocoagulation was first investigated in the 1950s by Meyer-Schwickerath as an approach to halt advancement of retinal detachment, cf.\ \citep{Meyer-Schwickerath1954}. Originally, he focused sunlight on the retina to induce spatially confined lesions due to the heating of the irradiated tissue above the damage threshold (coagulation). Modern sophisticated medical setups for retinal photocoagulation typically comprise pulsed laser exposure in the range of $20\,$-$\,400\, \text{ms}$ with lasers in the green spectral range. Today, this technique has become a clinical standard, whose scope of treatment extends also to several other retinal diseases such as diabetic retinopathy and macula edema, cf.\ \citep{mmETDRSRG.1991,mmETDRSRG.1985}, or ischemia due to vein occlusion, cf.\ \citep{mmBVOSG.1986, mmTCVOSG.1997,mmShah.2011}. The key element for a successful therapy is the correct dosage of laser radiation. While some diseases, such as retinal detachment, require high laser energy to induce significant coagulation, other diseases, such as chronic central serous retinopathy, only require insignificant heating in order to stimulate intercellular processes without inducing damage, cf.~\citep{mmLavinsky.2015}.

A central factor in photocoagulation is the absorption coefficient of the retina. This parameter varies strongly from patient to patient and even spatially over a single retina. This poses a major challenge in retinal laser therapy as the absorption coefficient is unknown and therefore also the appropriate laser power for a safe and effective treatment. In case of manual treatment, this is highly dependent on the physician's experience. In order to overcome these drawbacks, \citep{Brinkmann.2012} have developed an approach to measure the temperature increase during photocoagulation by means of the photoacoustic interaction of light and biological tissue.  
Briefly, if a short light pulse with a duration in the range of $100\, \text{ns}$ is absorbed, the irradiated tissue undergoes thermoelastic expansion which gives rise to a pressure transient. This pressure transient can be measured by means of piezoelectric transducers attached to the surface of the eye globe. By using the temperature dependence of the Grüneisen coefficient, temperature increase of the probed volume can be computed from the pressure transients. However, this averaged and depth-weighted volume temperature is less important for control as the peak temperature in the irradiated volume is crucial for the success of the treatment. To this end, in \citep{Baade.2013}, an approximation of the underlying heat diffusion is calculated and employed for control. Peak-temperature control based on the approximate conversion from peak to volume temperature has been demonstrated in open-loop and closed-loop experiments in  \citep{Baade.2017} and \citep{Herzog.2018}. 
With the overall goal of increasing the safety, accuracy, and reliability of peak temperature control, a method for real-time estimation of the absorption coefficient based on a discrete, reduced-order model of the heat diffusion equation was developed in \citep{Kleyman2020}. 
To this end, the authors generalized the parametric model order reduction (pMOR) proposed in \citep{Baur11}.
In the previous work \citep{Kleyman2020}, the parameter estimation was concluded via measurements of the peak temperature, which can, in general, be hard to obtain. A particular novelty of the present work is that we use a volume temperature for state and parameter estimation, which is more accessible in terms of our application. Further, compared to the previous work, we present results also for experimental data obtained from porcine eyes.

The main contribution of this paper is the development of tailored state estimation and parameter identification  
based on the measured volume temperature. The states are modeled by a linear system while the parameter dependence is nonlinear. Henceforth, one can express the peak temperature as a function of estimated states and the parameter, i.e., as another output of the system. 
In particular, we begin with an extension of the model and pMOR presented in \citep{Kleyman2020}. To this end, we employ the modeling of the optoacoustically determined volume temperature as proposed in \citep{Brinkmann.2012} for pMOR with polynomial parameter dependencies in the input and output operator. 
We utilize the obtained discrete-time, but parameter-dependent, state-space model for the design of an extended Kalman filter (EKF) and a moving horizon estimator (MHE). To encompass the parametric dependence, we extend the state-space model to allow estimation of the parameter in addition to the states. 
We compare both estimators and show results on simulated and experimental data.    

The remainder of this paper is structured as follows. 
In Section~\ref{sec:setup} we present modified experimental setup using one laser only. In Section \ref{sec:modeling} the modeling of the heat diffusion relating to our setup as well as the spacial discretization and parametric MOR are introduced. Section~\ref{sec:state_param_est} provides the state and parameter estimation where we employ an extended Kalman filter and a moving horizon estimator, followed by a comparison of both. Section \ref{sec:conclusion} concludes with a summary and directions for future work.

%
\section{Experimental setup} \label{sec:setup}

In earlier works regarding temperature-controlled laser therapy (\citep{Brinkmann.2012, Schlott.2012, Herzog.2018}) two lasers, a cw laser for heating and a pulsed laser for temperature probing, have been collinearly superimposed. We further developed the setup to one pulsed laser which can be used for both, heating and measuring. 

\begin{figure}
	\centering
	\includegraphics[width = \columnwidth]{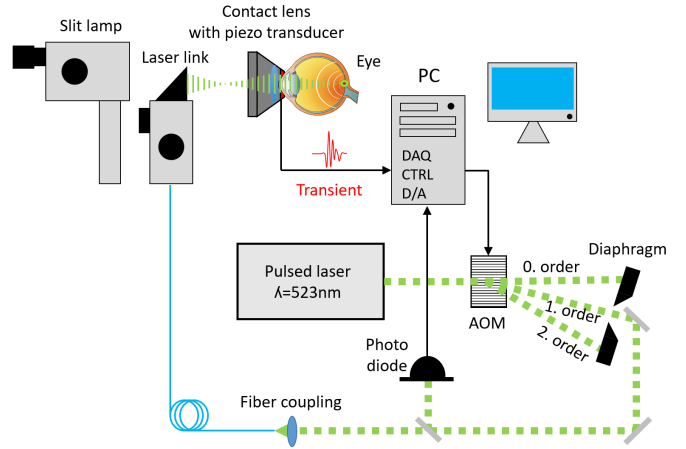}
	\caption{Schematic sketch of the experimental setup.}
	\label{fig:Setup_01}
\end{figure}

Figure \ref{fig:Setup_01} shows a simplified scheme of the experimental setup. A pulsed laser beam in the green spectral range (\mmVal{\lambda}{523}{nm}) is guided through an acousto-optic modulator (AOM). An AOM is an acousto-optic device which comprises a piezo transducer which is coupled to a crystal. By applying an AC voltage to the transducer a grating is induced in the crystal which allows for diffracting the incident beam in several orders of diffraction. The zero order is passed through the crystal collinearly to the incident beam, higher orders of diffraction are deflected as shown for the first and second order. The amount of energy deposited to the single orders of diffraction can be controlled via the amplitude of the AC voltage as indicated by the arrow pointing on the AOM. The first order of diffraction is passed through a diaphragm, other orders of diffraction are blocked. A small portion of the passed beam is deflected to a photo diode in order to normalize the pressure transients to the laser pulse energy applied. Afterwards, the passed beam is coupled to an optical fiber which is connected to the laser link of a slit lamp. A slit lamp is an ophthalmological instrument which is used by ophthalmologists to examine the exterior part of the eye. If the eye’s background (fundus) is to be examined or treated by means of laser radiation, additional optics such as contact glasses are needed. In order to enable acoustic measurements, a commercial contact glass was equipped with a ring-shaped piezo transducer.
The setup is operated by a personal computer equipped with high speed data acquisition and D/A boards. Currently, the laser is operated with a repetition rate \mmVal{f_{rep}}{10}{kHz}. Each 40th laser pulse is used to measure the temperature at a certain, calibrated energy $E_{cal}$. For the following 39 laser pulses the pulse energy is set to a heating energy $E_{heat}$. This yields a measurement rate of \mmVal{f_{meas}}{250}{Hz}.

\section{Modeling of heat diffusion}\label{sec:modeling}

\begin{figure}
	\centering  
	\includegraphics[width = \columnwidth]{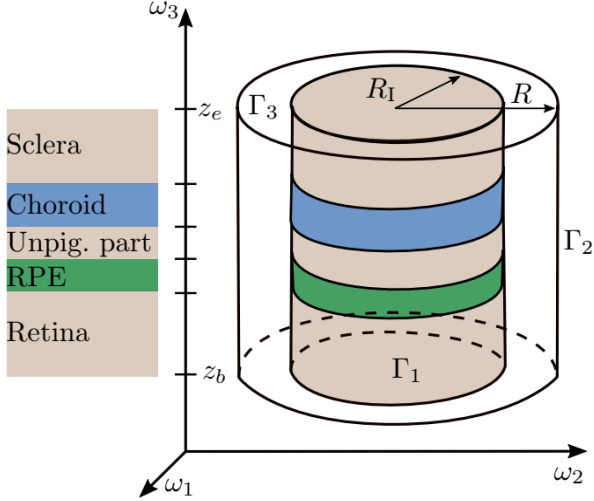}
	\caption{Schematic illustration of the five considered layers of the eye fundus and the cylinders. Figure adapted from \citep{Kleyman2020}.}
	\label{fig:cylinders}
\end{figure}

Having presented the setup in the laboratory, we will now present the model we use for estimation and control.
We model heat diffusion inside the tissue by a linear parabolic partial differential equation (PDE) and present a parametric model order reduction (pMOR).  
For a more detailed explanation we refer to \citep{Kleyman2020}. \\
We consider five different tissue layers of the eye fundus for modeling: retina, retinal pigment epithelium (RPE), an unpigmented part of the RPE/choroid, choroid and sclera as shown in Fig.~\ref{fig:cylinders}. As the experiments are carried out on porcine eyes, we consider average values for the thickness $d$ of each layer from \citep{Brinkmann.2012} as shown in Tab. \ref{tab:absorption}. The absorption in these layers varies strongly and most of the light is absorbed in the RPE and choroid. Therefore, we neglect any absorption in the retina, sclera and the unpigmented part.
The laser irradiates a round spot, where the intensity of the radiation decreases in depth due to the absorption of light according to the Lambert-Beer law. Since the irradiated volume can be modeled as a cylinder, the surrounding tissue is also modeled as a (finite) cylinder. Thus, the considered volume consists of an inner (irradiated) cylinder and a larger, outer cylinder. At the boundaries $\Gamma:=\Gamma_1\cup\Gamma_2\cup\Gamma_3$ of the outer cylinder, we assume Dirichlet boundary conditions. This is admissible as long as the outer cylinder is chosen sufficiently large. The heat diffusion can be described in the domain $\Omega\subset\mathbb{R}^3$ of the outer cylinder by 
\begin{align}
\rho C_p \frac{\partial T(\omega,t)}{\partial t}-k\Delta T(\omega,t)=Q(\omega,t)\; \forall\,  
(\omega,t)\in\Omega\times(0,t_\mrm{f})
\label{eq:pde}
\end{align}
with boundary and initial conditions 
\begin{align}
\begin{split}
T(\omega,t)&=0\quad \text{$\forall\, (\omega,t)\in\Gamma\times (0,t_\mrm{f})$},\\ 
T(\omega,0)&=0 \quad \forall\, \omega\in\Omega.
\end{split}
\label{eq:BC}
\end{align}
Here, $T(\omega,t)$ describes the temperature difference between the ambient space and the tissue. Thus, the initial temperature difference is zero. The heat capacity $C_p$, the thermal conductivity $k$ and the density $\rho$ are assumed to be constant and the same to those of water (${\rho = 993\text{ kg/m}^3}$, $C_p = 4176\text{ J/(kgK)}$, ${k = 0.627\text{ W/mK}}$), the main component of tissue, cf.\ \citep{Baade.2017}.
\begin{table*}[t]
	\centering
	\caption{Porcine eye fundus: average thicknesses and absorption coefficients from \citep{Brinkmann.2012}}
	\begin{tabular}{l|c|c|}
		Layer & Thickness ($10^{-6}\,\mrm{m}$) & Absorption coefficient ($10^2\,\text{m}^{-1}$)\\
		\hline
		Sclera & $d_\mrm{sc} = 139\;\,$ & $0$\\
		Absorbing part of choroid & $d_\mrm{ch}=400\quad$ & $\mu_\mrm{ch} =270$\\
		Unpigmented part of RPE/choroid & $d_\mrm{up}=4\quad\,$ & $0$ \\
		RPE & $\,d_\mrm{rpe} =6\qquad$ & $\mu_\mrm{rpe} = 1204$ \\
		Retina & $\,d_\mrm{r} = 190\,$ & $0$\\	
	\end{tabular}
	\label{tab:absorption}
\end{table*}
The light-tissue interaction is modeled as a heat source $Q(\omega,t)$ on the right-hand side of \eqref{eq:pde} and is given by Lambert-Beer law:
\begin{align}
\label{eq:heat_source}
Q(\omega,t):=
\begin{cases}\frac{u(t)}{\pi R_\mrm{I}^2}\mu(\omega_3) e^{-\int_0^{\omega_3}\mu(\zeta)\mrm{d}\,\zeta}, &\text{if}\, \omega_1^2+\omega_2^2\leq R_\mrm{I}^2,\\ 
0,&\text{otherwise,}\end{cases}
\end{align}
where $u : [0,T]\to \mathbb{R}_{\geq 0}$ is the laser power, ${R_I = 1\cdot 10^{-4}\, \text{m}} $ the radius of the irradiated spot and $\mu \in L_\infty(\Omega,\mathbb{R}_{\geq 0})$ is an absorption coefficient. In particular, this parameter is unknown and can change from treatment spot to treatment spot. 
Hence, we explicitly denote the dependence of the unknown parameter $\mu(\omega)\equiv \mu(\omega_3)$ for which we will (after discretization) carry out a particularly suited parametric model order reduction in Subsection~\ref{subsec:pMOR} to perform parameter estimation in Section~\ref{sec:state_param_est}. In our particular application it has shown that absorption only takes place in the RPE and choroid, hence
\begin{align*}
\mu(\omega) = \begin{cases}
\mu_{rpe}, \quad &\text{if } \omega_3 \in [z_b+d_r,z_b+d_r+d_{rpe}],\\
\mu_{ch}, \quad &\text{if } \omega_3 \in [z_b+d_r+d_{rpe}+d_{up},z_e-d_{sc}],\\
0, &\text{otherwise},
\end{cases}
\end{align*}
where $z_b$ and $z_e$ are defined in Fig.\ \ref{fig:cylinders}.

The parabolic PDE \eqref{eq:pde} can be restated as a linear state-space model in the Hilbert space $L_2(\Omega)$, i.e., 
\begin{align}
\label{eq:statespacemodel}
\frac{\partial x(t)}{\partial t}=\mathcal{A}x(t) + \mathcal{B}(\mu)u(t), \qquad x(0) = 0,
\end{align}
where $\mathcal{A}:D(\mathcal{A})\subset L_2(\Omega) \to L_2(\Omega)$ is the generator of a strongly continuous semigroup on $L_2(\Omega)$ and $\mathcal{B}\in L(\mathbb{R},L_2(\Omega))$ is a bounded control operator. More precisely, we set $\mathcal{A}=\Delta$ endowed with the domain $D(\mathcal{A}) = H^2(\Omega)\cap H^1_0(\Omega)$. In particular, \eqref{eq:statespacemodel} is a well-posed system in the sense that for any $u\in L_1(0,T;\mathbb{R})$ we obtain a unique solution $x\in C(0,T;L_2(\Omega))$. For details, we refer to, e.g., \citep[Section 2]{Curtain1995}.
%
%
\subsection{Volume Temperature} \label{subsec:vol_temp}
Having defined the state space model in \eqref{eq:statespacemodel}, we will now define the output operator that will model the system's output. We emphasize that in our case we have to consider two outputs: on one hand the volume temperature that will represent the measurements, and on the other hand the peak temperature that is to be controlled for a successful treatment. We consider the volume temperature as the system's output although, strictly speaking, we actually measure pressure, cf. Sec.~\ref{sec:setup}. The following section concisely describes the modeling of the volume temperature as carried out in \citep{Brinkmann.2012}. For clarity of presentation we will use cylindrical coordinates $(r,\varphi,z)$.

First, we calculate the mean temperature $x_{\mrm{mean}}$ of the irradiated area in each layer $z$, where the intensity of the laser light is (assumed to be) constant, i.e.
\begin{align*}
x_\mrm{mean}(t,z) &=\frac{1}{\pi R_\mrm{I}^2}\int_{0}^{2\pi} \mrm{d}\,\phi \int_{0}^{R_\mrm{I}} r x(r,z,t)\mrm{d}\,r.
\end{align*}  
The volume temperature can then be expressed by the integral over all temperatures, weighted by the absorbed laser power at $z$
\begin{align}
x_\mrm{vol}(t)= \int_{z_b}^{z_e} x_\mrm{mean}(t,z) \mu(z)e^{\int_{0}^{z}\mu(\zeta)\diff \zeta} \mrm{d}z,
\label{eq:Tvol}
\end{align}
where $z_e-z_b$ is the length of the cylinder. 
Considering absorption in the RPE and choroid yields 
\begin{align*}
\begin{split}
x_\mrm{vol}(t) = &\int_{0}^{d_{rpe}}x_\mrm{mean}(t,z)\mu_\mrm{rpe}e^{-\mu_\mrm{rpe}z}\diff z+\\ &\int_{d_b}^{d_e}x_\mrm{mean}(t,z)\mu_\mrm{ch}e^{-\mu_\mrm{rpe} d_\mrm{rpe}-\mu_\mrm{ch}(z-d_b)}\diff z
\end{split}
\end{align*}
with $d_b = d_{rpe}+d_{up}$ and $d_e = d_{rpe}+d_{up}+d_{ch}$.
The output operator depends, similar to the input operator in \eqref{eq:statespacemodel}, on the absorption coefficient $\mu$.

Hence, we define an output operator $\mathcal{C}(\mu)\in L(L_2(\Omega),\mathbb{R})$ via 
\begin{align*}
\mathcal{C}_\text{vol}(\mu)x := \int_{z_b}^{z_e} x_\mrm{mean}(t,\omega_3) \mu(\omega_3)e^{\int_{0}^{\omega_3}\mu(\zeta)\diff \zeta} \mrm{d}\omega_3.
\end{align*}
The aforementioned volume temperature can be obtained from the measurements and is central in the state and parameter estimation. For control, however, the peak temperature is the decisive quantity that determines success of the treatment.
To this end, we extend the output by the peak temperature before the model order reduction (MOR) to obtain a reduced order model that is well-suited for both estimation and control. 
Numerical simulations suggests that the peak temperature occurs in the center of the RPE layer during heating. For this reason, we extend the output operator to  
\begin{align*}
\mathcal{C}:=\begin{pmatrix}
\mathcal{C}_\text{vol}(\mu)\\
\mathcal{C}_\text{peak}
\end{pmatrix}
\end{align*}
with 
\begin{align*}
(\mathcal{C}_\text{peak}x)(\omega):= \begin{cases}
x(\omega)  &\text{for } \omega_1=\omega_2=0, \omega_3= \frac{d_{rpe}}{2}\\
0  &\text{otherwise.}
\end{cases}
\end{align*}
We note that strictly speaking, this operator is not bounded, i.e. $\mathcal{C}_\text{peak}\notin L(L_2(\Omega),\mathbb{R})$. However, it can be shown by classical PDE methods, cf.\ cf.\ \citep{Evans2010}, that
the solutions of \eqref{eq:pde} enjoy a higher spatial regularity due to the smoothness of the coefficients on the subdomains and the finite dimensional control such that $x(t)\in C(\Omega,\mathbb{R})$ and hence a point evaluation makes sense. Due to space limitations, we will not go into detail here.

The output we consider in the subsequent sections is then given by
\begin{align}
y(t) = \mathcal{C}x(t). 
\end{align}
%
%
\subsection{Polynomial Approximation and Discretization}\label{subsec:taylor_disc}
In this section, we prepare for the parametric model order reduction by Taylor approximation and spatial discretization. 
Hence, we will first perform a Taylor series approximation of the input and output operator via
\begin{align*}
\mathcal{B}(\mu)&\approx  \sum_{i = 0}^{k_b} \frac{\partial^i \mathcal{B}}{\partial \mu^i}(\mu_0)\frac{(\mu-\mu_0)^i}{i!} \\ 
\mathcal{C}(\mu) &\approx \sum_{i = 0}^{k_c} \frac{\partial^i \mathcal{C}}{\partial \mu^i}(\mu_0)\frac{(\mu-\mu_0)^i}{i!} = \sum_{i = 0}^{k_c} \frac{\partial^i \mathcal{C}}{\partial \mu^i}(\mu_0)\frac{(\alpha \mu_0)^i}{i!},
\end{align*}
where $\mu(\omega) = (\alpha+1)\mu_0(\omega_3) $ and $\mu_0$ is the one of mean absorption coefficients listed in the right column of Tab. \ref{tab:absorption}, depending on $\omega_3$. Hence, in the following, we parameterize $\mu$ by the scalar prefactor $\alpha \in \Dc$, where $\Dc$ is the parameter domain.

In the next step, we perform a spatial discretization via finite differences with $n_f\in \mathbb{N}$ discretization points, cf.\ \citep{Kleyman2020}, and obtain the finite dimensional state space model
\begin{align}
\label{eq:fullsys}
\begin{split}
\dot x(t)&=Ax(t)+b(\alpha)u(t),\quad x(0)=0\\\quad y(t)&=C(\alpha)x(t),\quad t\geq 0.
\end{split}
\end{align}
where $A\in \R^{n_f \times n_f}$, $x \in \R^{n_f}$, $y\in \R^{2}$ and $b$ and $C$ are polynomials of the form
\begin{align}
\label{eq:bann}
b(\alpha)=b_0+\sum_{i=1}^{k_B}\alpha^ib_i,\quad C(\alpha)=c_0^\top+\sum_{i=1}^{k_C}\alpha^ic_i^\top,\quad 
\end{align}
where $b_i\in\mathbb{R}^{n_f}$, $i=0,\ldots,k_B$, $c_i\in\mathbb{R}^{2\times n_f}$, $i=0,\ldots,k_C$ and $n_f \in \mathbb{N}$ is the number of discretization points. Due to the rotational symmetry of the irradiated area, the discretization is carried out in cylindrical coordinates in the $r\,z$-plane. 
%
\subsection{Parametric Model Order Reduction}\label{subsec:pMOR}
In order to enable real time (optimal) control of the high-dimensional system \eqref{eq:fullsys}, we apply model order reduction (MOR). In that context, we have to keep the parameter dependence. To this end, we generalize the parametric model order reduction (pMOR) in \citep{Baur11}. This approach was already successfully applied in the previous work \citep{Kleyman2020}. We showed that the $\Hc_2$-optimal reduction of the parameter dependent transfer function over $L^2(\Dc)\otimes \Hc_2$ is the same as the optimal reduction of a parameter independent transfer function with respect to the $\Hc_2$-norm. 

After the pMOR, we obtain the reduced order model of order $n$ with the global basis $W^\top \in \R^{n \times n_f}$ and $V \in \R^{n_f \times n}$
\begin{align*}
\begin{split}
W^\top V \dot{x}_r(t) &= W^\top A V x_r(t)+ W^\top b(\alpha)u \\
y_{r}(t)&=C(\mu)V x_r(t),
\end{split}
\end{align*}
which is
\begin{align}
\begin{split}
\dot x_r(t)&=A_rx_r(t)+b_\mrm{r}(\alpha)u(t),\\
y_{r}(t)&=C_r(\alpha)x_r(t),
\end{split}
\end{align}
with $A_r = (W^\top V)^{-1}W^\top AV$, $b_r(\alpha) = (W^\top V)^{-1}W^\top b(\alpha)$ and $C_r(\alpha) = C(\alpha)V$.\\

\section{state and parameter estimation}
\label{sec:state_param_est}
As described in Sec. \ref{subsec:vol_temp}, only the volume temperature can be measured. However, the peak temperature needs to be controlled to avoid undesired damage at the irradiated spot. Therefore, it is necessary to estimate the states of the reduced model as well as the unknown absorption coefficient. We consider two different methods for state and parameter estimation: an extended Kalman filter and a moving horizon estimator. In both methods, we do not estimate the absorption coefficient directly, but rather the prefactor $\alpha$. The absorption coefficient can then be obtained via the relation $\mu=(\alpha+1)\mu_0$. 
In the following, we consider the reduced-order, discrete-time state space model
\begin{align}
\label{eq:reduced.parametric.model}
\begin{split}
x_{k+1} &= A_\mrm{d} x_k + b_\mrm{d} (\alpha) u_k \\ 
y_k &= c_{\mrm{d,vol}}(\alpha_k) x_k.
\end{split}
\end{align} 
We consider a sampling rate of $250\, \text{Hz}$ for simulations and experiments.  
\subsection{Extended Kalman Filter}\label{sec:EKF}
The extended Kalman filter (EKF) is a well known state estimator for nonlinear systems. It is based on the linearization of a nonlinear model subject to process noise $w$ and measurement noise $v$ which are assumed to be uncorrelated and normally distributed, see e.g. \citep{Chui2017}. In our application, the EKF can not only be used for state but also for parameter estimation by extending the state by $\alpha$, i.e. 
\begin{align}
\begin{split}
\bar{x}_{k+1} &= 
\begin{pmatrix}
x_{k+1} \\ \alpha_{k+1} \end{pmatrix} = f(x_k,\alpha_k,u_k,w_k) \\&= \begin{pmatrix}
A_\mrm{d} & 0\\ 0 &1 
\end{pmatrix} \bar{x}_k + \begin{pmatrix}
b_\mrm{d}(\alpha_k)\\ 0
\end{pmatrix} u_k +w_k, \\ 
y_k &= g(x_k,\alpha_k)+v_k = \begin{pmatrix}
c_{\mrm{d,vol}}(\alpha_k) & 0
\end{pmatrix} \bar{x}_k + v_k.
\end{split}
\label{eq:ss_EKF}
\end{align}

The EKF algorithm can be divided into two steps. First, the a priori system state $\bar{x}_k^-$ and estimation error covariance matrix $P_k^-$ are calculated from the previous estimates
\begin{align}
\begin{split} 
\bar{x}_k^-&=f(\bar{x}_{k-1},u_{k-1})\\
P_k^-&=A_{k-1}P_{k-1} A_{k-1}^\top+Q,
\end{split}
\end{align}
where $Q \in \R^{n+1\times n+1}$ is a process covariance matrix and $A_{k-1} \in \R^{n+1\times n+1}$ is the Jacobian of $f(\bar{x}_{k-1},u_{k-1})$. For the nominal system of \eqref{eq:ss_EKF}, this Jacobian is given by
\begin{align*}
A_{k-1} &= \begin{pmatrix}
A_\mrm{d} & \frac{\partial f(\bar{x}_{k-1},u_{k-1})}{\partial \alpha_{k-1}}|_{\bar{x}_{k-1},u_{k-1}}\\ 0 &1 
\end{pmatrix}.
\end{align*}
Second, the estimation error covariance matrix $P_k$, the Kalman gain $H_k \in \R^{1 \times n+1}$ and the estimated state $\bar{x}_k$ are calculated as
\begin{align}
\begin{split}
H_k &= P_k^- c_k^\top (c_k P_k^- c_k^\top+R)^{-1}\\
\bar{x}_k &= \bar{x}_k^- +H_k (y_k- g(\bar{x}_k^-))\\
P_k&= (I_{n+1}-H_k c_k)P_k^-
\end{split}
\end{align}
with the identity $I_{n+1} \in \R^{n+1 \times n+1}$, the measurement covariance $R$ and the Jacobian of the output ${c_k = \frac{\partial g(\bar{x}_k)}{\partial \bar{x}_k}|_{\bar{x}_k^-}}$. The matrices $Q$ and $R$ are design parameters that weight the reliability of the model and the measurement. 
Note that the pair $(A_{k-1},c_k)$ is not observable if ${u=0}$. However, as the system is heated while estimating this is not a problem for our application. 

As the states are several orders of magnitude smaller than the output and the prefactor $\alpha$, a similarity transformation with the transformation $T = \mrm{diag}(10^{-8},10^{-8},10^{-8},1) $ is performed. This allows for a more intuitive scaling of $Q$ and $R$.

\subsection{Moving Horizon Estimation}
In alignment with the EKF, moving horizon estimation (MHE) is another state estimation strategy, which affords a wide range of online applications in nonlinear processes \citep[Chapter 4]{Rawlings.2017}.

MHE is an optimization based approach for state estimation that uses a sequence of $N$ most recent measurements to estimate the current state at time $T$. In our present setting, we define and apply our MHE algorithm to the reduced model in \eqref{eq:ss_EKF} as follows. First, we denote $\Vert v\Vert_{M^{-1}} := v^TM^{-1}v$ for a vector $v\in \mathbb{R}^k$ and matrix $M\in \mathbb{R}^{k\times k}$ with suitable dimension $k\in \mathbb{N}$.

Given $N$ most recent measurements corresponding to a control sequence $\left(u_{T-N},u_{T-N+1},\ldots,u_T\right)\in \mathbb{R}^{N+1}$ and a guess for the initial state and parameter $\chi_0\in \mathbb{R}^{n+1}$, the estimated state trajectory ${\bf x}=\left(x_{T-N},\ldots, x_T\right)\in \mathbb{R}^{(N+1)n}$ and parameters $\boldsymbol{\alpha}=\left(\alpha_{T-N},\ldots,\alpha_T\right)\in \mathbb{R}^{N+1}$ at the time $T$ are given as the solution to the following optimization problem:
\begin{align}
\label{problem:mhe.optimization}
\min_{{\bf x}\in \mathbb{R}^{(N+1)n},{\boldsymbol{\alpha}} \in \mathbb{R}^{N+1}}
J({\bf x},\boldsymbol{\alpha})
\quad \text{s.t. } \boldsymbol{\alpha} \in \mathcal{D}^{N+1},
\end{align}
where $J$ is defined to be
\begin{align*}
&J(\bf{x},\boldsymbol{\alpha}):= \\
&\left\| \begin{pmatrix}
x_{T-N}\\
\alpha_{T-N}
\end{pmatrix}-\chi_0\right\|^2 _{P^{-1}} +  \sum_{k=T-N}^{T} \vert y_k - c_{\mrm{d,vol}}(\alpha_k) x_k\vert^2_{R^{-1}}\\
&+ \sum_{k=T-N}^{T-1} 
\left\| \begin{pmatrix}
x_{k+1} - A_\mrm{d} x_k + b_\mrm{d} (\alpha_k) u_k \\
\alpha_{k+1} - \alpha_k
\end{pmatrix}\right\|^2_{Q^{-1}},
\end{align*}
with symmetric positive definite $P,Q\in\R^{{n+1}\times {n+1}}$ and $R\in\R_{>0}$. These quantities serve, together with the horizon $N$, as design parameters which we will specify later.

Upon solving \eqref{problem:mhe.optimization}, $x_T$ and ${\alpha}_T$ are taken to be the current estimate for the state and parameter and we update the initial guess for the state and parameter via 
$\chi_0=\left(\begin{smallmatrix}
x_{T-N+1}\\\alpha_{T-N+1}
\end{smallmatrix}\right)$. Subsequently, we set $T=T+1$ and reiterate the procedure. We emphasize that in every iteration a trajectory of length $N+1$ is computed by solving the optimization problem while only the current state and parameter at time $T$ serve as estimators. Whenever $T<N$, we set $N=T$, i.e., we use only the available amount of measurements to solve the optimization problem in \eqref{problem:mhe.optimization} to estimate the current state and parameter.
 
\subsection{Results}

\begin{figure}
	\centering
	\includegraphics{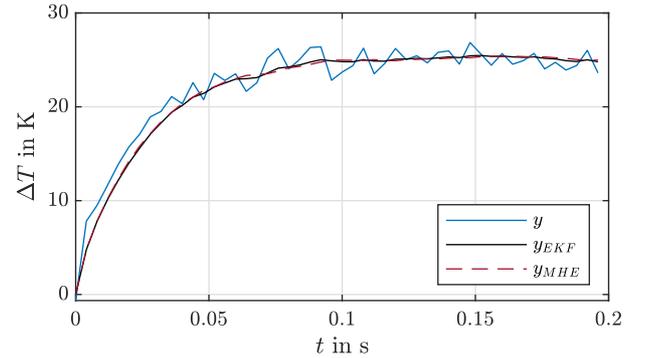}
	\caption{Simulated volume temperature with noise in blue and estimated output in black (EKF) and in dashed, red (MHE).}
	\label{fig:y_EKF}
\end{figure}

In this section, we present first results of the two estimating strategies, EKF and MHE.
Fig. \ref{fig:y_EKF} shows the estimated output of the EKF and MHE and the system's output for simulated data. We choose the initial values $x_0=0$, see \eqref{eq:BC}, $\alpha_0=0$, the constant input signal $u\equiv 30\, \mrm{mW}$ and we add Gaussian white noise with a variance of $1\, \mrm{K}$ to the output. We consider a ROM of third order and $k_b = 8$.
In order to compare the results obtained in MHE with EKF in a consistent manner, we choose the corresponding design parameters equal, i.e., $Q = \mrm{diag}(10^{-3}, 10^{-3},10^{-3},0.15)$ and $R=10^{2}$ for both EKF and MHE. For the arrival cost and horizon in the MHE implementation, we choose $P=Q$ and $N=5$.

\begin{figure}
	\centering
	\includegraphics{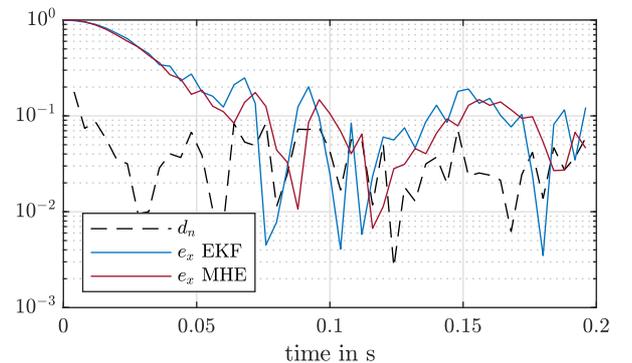}
	\caption{Comparison of the relative estimation error $e_x$ (including parameter) in blue (EKF) and red (MHE) and relative noise $d_n$ in black (dashed line).}
	\label{fig:error_EKF}
\end{figure}

Fig. \ref{fig:error_EKF} shows the relative error $e_{x}(t) = \frac{||x(t)-x_\mrm{EKF,MHE}(t)||}{||x(t)||}$ of the state and parameter estimation over time for simulated data as well as the "relative noise" between the simulated output without and with noise $d_n(t) =\frac{||y(t)-y_{noise}(t)||}{||y(t)||}$. The relative estimation error shows that the states converge to a neighborhood of the actual values for MHE and EKF. 
A more accurate estimate of the state is not to be expected due to the noise as $d_n(t)$ is of the same order of magnitude. The error when using MHE is similar to the error for EKF before approaching the magnitude of the relative noise.

\begin{figure}
	\centering
	\includegraphics{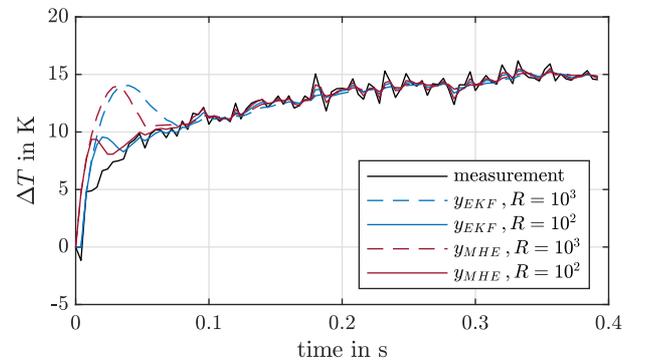}
	\caption{Measured volume temperature in black and estimated output with $R = 10^3$ (dashed) and with $R = 10^2$ (solid). MHE estimates are pictured in red, EKF estimates in blue.}
	\label{fig:y_meas_EKF}
\end{figure}

\begin{figure}
	\centering
	\includegraphics{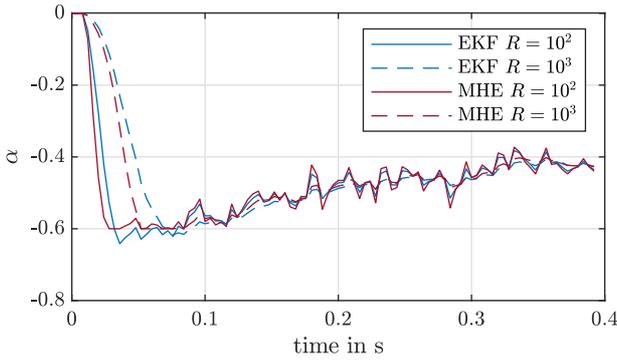}
	\caption{Estimated prefactor $\alpha$ of the absorption coefficient with $R = 10^3$ (dashed) and with $R = 10^3$ (solid). MHE estimates are pictured in red, EKF estimates in blue.}
	\label{fig:alpha_meas_EKF}
\end{figure}

Experimental results of the measured and estimated volume temperature are shown in Fig.~\ref{fig:y_meas_EKF}. We tested our implementation on the measured volume temperature with covariances $R = 10^3$ and $R = 10^2$ for both estimators. 
The estimation of the output, especially in the first $0.1\, \text{s}$, depends very much on the weighting $R$. The state converges faster for smaller values of R but one is also more prone to overfitting. This is also illustrated in Fig.~\ref{fig:alpha_meas_EKF}, where the estimation of the parameter $\alpha$ is shown.
The initial overshooting in the estimations in the beginning of the heating process appears to be of similar magnitude for MHE and EKF. However, the MHE converges faster than the EKF algorithm. After the $0.1\, \text{s}$ mark, they appear to equally well-estimate the measured temperature.

Again, a slight outperformance of MHE over EKF is seen in Fig. \ref{fig:alpha_meas_EKF}. In this figure, MHE with respective choices of $R$ starts estimating the value of $\alpha$ slightly sooner than the EKF with the corresponding $R$ with relates to a shorter duration of overshoot in the estimation of the volume temperature.

\begin{figure}
	\centering
	\includegraphics{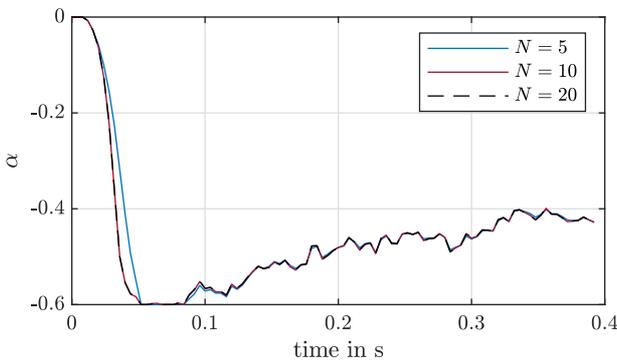}
	\caption{Estimated prefactor $\alpha$ of the absorption coefficient for different horizon $N$ and $R=10^3$.}
	\label{fig:alpha_compare_N}
\end{figure}

In the end, as it is natural to suspect that increasing the horizon leads to a better estimation, we compared the performance of MHE for different horizon length. We compared the estimations of the absorption coefficient $\alpha$ with respect to the following horizon length $N=5,~10,~20$ in Fig. \ref{fig:alpha_compare_N}.
In this plot, $N=10$ clearly outperforms $N=5$, though $N=20$ is hardly distinguishable from $N=10$.

Both the MHE and the EKF implementations proved to be effective in estimating the absorption coefficient and the measurements. 
Although MHE seems to slightly outperform EKF in both the estimated measurement and the parameter identification, it is computational more expensive than the EKF. On the other hand, an advantage of the MHE is that we can consider constraints on the absorption coefficient, i.e. $\boldsymbol{\alpha} \in \Dc^{N+1}$. These first results motivate further investigation with respect to particular tuning of both EKF and MHE (i.e., e.g., arrival cost and weighting matrices) and with respect to a wider range of experimental data.

\section{Conclusion}\label{sec:conclusion}
In this paper, we have presented ways for parameter and state estimation in retinal laser therapies. For this purpose, we have extended the modeling of heat diffusion to include the volume temperature, which is modeled as a system output. We reduced the high-dimensional system using pMOR to obtain a model that is real-time capable and additionally allows for the estimation of the absorption coefficient. We applied two methods for parameter and state estimation, EKF and MHE, and compared the results both in simulation and with measured data. Both estimators performed similarly well. The MHE converges faster, but this is accompanied by a considerably higher computational effort.
To further improve the estimation, adaptive adjustment of the weighting matrices might allow faster convergence and yet smoother estimation of the states. 
The application of model predictive temperature control and further investigation of the estimators are part of future work.

\bibliography{references}           
\end{document}